# Superdeformed bands in $^{32}$S and neighboring nuclei predicted within the Hartree-Fock method

H. Molique,[1,2] J. Dobaczewski,[1,3] and J. Dudek[1]

[1]*Institut de Recherches Subatomiques, CNRS-IN$_2$P$_3$/Université Louis Pasteur, F-67037 Strasbourg Cedex 2, France*

[2]*Institut Universitaire de Formation des Maîtres d'Alsace, F-67100 Strasbourg, France*

[3]*Institute of Theoretical Physics, Warsaw University, Hoża 69, PL-00681, Warsaw, Poland*

## Abstract

Superdeformed configurations in $^{32}$S, and in neighboring nuclei $^{33}$S, $^{31}$S, $^{33}$Cl, and $^{31}$P, are determined within the Hartree-Fock approach with the Skyrme interaction. Energies, angular momenta, quadrupole moments, particle-emission $Q$-values, and relative alignments and quadrupole moments are calculated for a number of superdeformed rotational bands in these nuclei. A new mechanism implying an existence of signature-separated rotational bands, distinct from the well-known signature-split bands, is discussed and associated with the time-odd channels of effective interactions.

Typeset using REVTEX

# I. INTRODUCTION

It is more than ten years by now that the study of superdeformed (SD) shapes in nuclei constitute one of the main venues in nuclear spectroscopy. Today it is well understood that an increased stability of strongly elongated nuclei results from the quantum (shell) effects that manifest themselves, among others, through a lowering of the nucleonic level densities at certain nucleon numbers. Within the anisotropic harmonic oscillator (HO) model, such shell effects arise when the ratios among the three principal frequencies are equal to the ratios of small integer numbers. The strongest shell closures correspond to axially deformed nuclei with the semi-axis ratios 2:1:1 and 3:2:2, the axis ratios being simply related to the oscillator frequencies [1].

The HO model is of course only poor an approximation for the majority of nuclei, for which the spin-orbit interactions play a determining role. Yet as it happens, the nuclear mean field obeys approximately a specific SU(2) symmetry, usually referred to as a pseudo-spin. (For an early formulation of the pseudo-spin symmetry see Refs. [2,3]; the contemporary formulation of the problem is based on the Dirac formalism according to the scheme proposed in Ref. [4] and further developed recently in [5,6].) Taking into account this symmetry allows at the same time to take care of the strong spin-orbit coupling, and profit from the simplicity of the HO model. Indeed, basing on the pseudo-spin symmetry, and employing a realistic deformed mean field Hamiltonian, it was possible to predict [7] (after the initial discovery of the SD band in $^{152}$Dy [8] but several years before the experiments in other regions have been done), the existence of the whole groups of SD nuclei. Moreover, the predictions gave also the fact that the deformations of strongly elongated shapes may considerably deviate from the 2:1:1 HO rule; these deviations are now confirmed through numerous experiments. The abundance scheme for the nuclear SD states at high angular momenta is well established today in the so-called $A{\simeq}190$, $A{\simeq}150$, $A{\simeq}130$, and $A{\simeq}80$ regions, see Refs. [9–12] for reviews; it includes also the recently discovered SD states in the $A{\simeq}60$ region [13–20], as well as a region of fission isomers in $A{\simeq}240$ nuclei, known already for a long time but at relatively low angular momenta.

Numerous cluster structures in light nuclei can also be interpreted as SD states, see Refs. [21,22] for more details and a more exhaustive reference list. In particular, the 2:1:1 deformed HO model predicts the SD shell-closures [1] at particle numbers 2, 4, 10, 16 and 28, a sequence characterized by an increased stability at large deformations, and also compatible with the $\alpha$-cluster structures. This gives, for example, the $\alpha$-$\alpha$ cluster ground state of $^{8}$Be, or the $^{16}$O-$\alpha$ cluster state in $^{20}$Ne. Prolonging the same sequence, one may expect stable SD structures in $^{26}$Ne and $^{32}$S. Next, doubly-magic SD states should appear at $N{=}Z{=}28$ (not to be confused with the spherical shell closures at the same nucleon numbers). However, because of the increasing role of the spin-orbit interaction when the nucleonic numbers increase, these values are slightly modified. This gives the doubly-magic SD nucleus $^{60}$Zn at the center of the experimentally known SD $A{\simeq}60$ region.

One can see that the SD states in $^{32}$S (although experimentally not discovered to date) constitute a missing link between the known cluster SD states in very light nuclei, and the known SD states in the $A{\simeq}60$ region. On the one hand, the first indications that the cluster SD states in $^{32}$S may exist are provided by the measurements in the $^{16}$O-$^{16}$O breakup channel [23], and by the $^{16}$O-$^{16}$O molecular resonances, as quoted in Ref. [24].

On the other hand, several mean-field calculations, both non-self-consistent [25] and self-consistent [26–29], as well as the α-cluster calculations [30,24], predict in $^{32}$S an existence of the 2:1:1 deformed structures. It is not clear at the moment, what exactly is the relationship between the molecular states (a pair of touching $^{16}$O nuclei), and the SD states (a compact matter distribution), although both classes may correspond to the same axis ratios and deformations. Such strongly deformed states should coexist with numerous low-deformation states already known in this nucleus [31]. In fact, the latter ones are very well described by the $sd$-shell model [32].

One may expect a number of interesting physical phenomena that can be studied in the hypothetical SD configurations of $^{32}$S, such as the shape coexistence, competition between various decay channels, proton neutron pairing and its deformation dependence, effects related to the time-odd components of nuclear mean fields, as well as nuclear-molecular and nuclear-cluster structures. Detailed properties may be significantly influenced by the presence of intruder states originating from the $N_0$=3, and even $N_0$=4 HO shells. With a total number of nucleons strongly restrained (only 16 per one kind of particles) one should expect a pronounced variation of shapes from one single-particle (particle-hole) configuration to another.

In the present paper we aim at theoretical description of the SD states in $^{32}$S and in four neighboring nuclei: $^{33}$S, $^{31}$S, $^{33}$Cl, and $^{31}$P. We present predictions pertaining to detailed spectroscopic information on excitation energies, spins, moments of inertia, and quadrupole moments of the SD rotational bands. All these observables may, in a very near future, become available within the discrete-spectroscopy measurements using large detector arrays; these observables have already been obtained experimentally in the other groups of SD nuclei.

The paper is organized as follows. After briefly presenting in Sec. II the theoretical methods we use in this study, in Sec. III we discuss the deformed-shell gaps and Coulomb effects in $^{32}$S, we present a classification of the SD bands, and describe the level crossings. Results of calculations for the SD bands in $^{32}$S are presented in Sec. IV, and those for $^{33}$S, $^{31}$S, $^{33}$Cl, and $^{31}$P in Sec. V. Finally, in Sec. VI we briefly discuss the question of the stability of SD bands, and Sec. VII presents our conclusions.

## II. THEORETICAL METHODS

In this paper we use the cranking Hartree-Fock (HF) method with the Skyrme SLy4 interaction [33]. The complete gauge-invariant [34] term $\boldsymbol{s}\cdot\boldsymbol{T}-\overleftrightarrow{J}^2$ has been removed from the Skyrme functional in order to comply with the procedure of adjusting the parameters of this force [33]. We solve the self-consistent HF equations by using the HFODD code (v1.75) [35,36], that employs the Cartesian HO basis. The basis used consists of the lowest $M$=306 HO states with the oscillator frequencies $\hbar\omega_z$=11.46 MeV and $\hbar\omega_\perp$=18.01 MeV. These parameters correspond to including in the basis up to $N_z$=14 and $N_\perp$=9 HO quanta. As discussed in Ref. [35], no further basis optimization is necessary, and thus the same unique basis has been used for all calculations presented below. In the calculations, the conservation of parity and signature symmetries has been assumed.

In Ref. [35] it was shown that by using a much larger HO basis of $M$=1200, one obtains a perfect agreement (up to 18 keV) of the $^{152}$Dy binding energies with those calculated using the space-coordinate code of Ref. [37]. At the same time, the $M$=300 calculations were giving a systematic underbinding of about 5 MeV, which, however, was very weakly dependent on the angular frequency or configuration. Here, we repeat similar tests in $^{32}$S. At the spherical shape, with $M$=306 we obtain the total energy of $E$=$-$270.000 MeV, while a simple one-dimensional coordinate-space code gives $E$=$-$270.876 MeV. At the $\hbar\omega$=1 MeV SD magic configuration of $^{32}$S (see below), the $M$=306 result for the total routhian is $R$=$E-\hbar\langle I_y\rangle$=$-$261.453 MeV, while $M$=1200 ($N_z$=24 and $N_\perp$=15) gives $R$=$-$262.124 MeV. From these results we conclude that the absolute energies of all nuclei presented in this paper should be shifted down, at all deformations and at all rotational frequencies, by a constant of about 0.8 MeV, in order to account for the finite size of the HO basis used in the calculations.

We have also performed the Hartree-Fock-Bogoliubov (HFB) calculations in $^{32}$S, by using the HFODD code (v1.79) for the zero-range (density independent) pairing interaction in the particle-particle channel, and the Skyrme SLy4 force in the particle-hole channel. The strength of the pairing interaction has been adjusted to obtain the value of the average pairing gap (at the spherical shape) equal to the one obtained with the help of the three-point mass staggering expression [38], applied to experimental masses of nuclei adjacent to $^{32}$S.

It turns out that the static pairing correlations, calculated within the HFB approximation for such a pairing strength, vanish at the SD shapes. With an artificially increased strength one may, of course, obtain non-zero pairing at the SD band-heads, but the HFB static pairing correlations disappear again very rapidly with increasing spin. Consequently, these calculations show that the proton-proton and the neutron-neutron pairing correlations do not contribute very strongly to the structure of the SD states in $^{32}$S, and may possibly affect the results only through dynamic correlations.

The above remarks do not exclude the possibility that, in the nuclei of interest here, strong proton-neutron pairing correlations may take place. To the contrary, in analogy to a suggestion relevant in the $A\simeq 60$ region [20], also in $^{32}$S we may expect strong proton-neutron pairing correlations at high spins. An approach which would take all these pairing mechanisms simultaneously into account is fairly complicated, and no appropriate tools exist to date to carry out such a program. Moreover, the experimental information about the proton-neutron pairing correlations at high spins should be considered as very limited today. Therefore the results presented below do not include the effect of the pairing interactions. This, as argued above, offers a reasonable approximation, and allows for a rapid overview of all available lowest-energy configurations. Such an analysis should be considered as a sufficient first step towards a more complete future study, given the fact that experimental results on the corresponding high spin effects do not exist at present.

## III. SINGLE-PARTICLE STRUCTURES AT $N=Z=16$

In this section we discuss the deformed-shell gaps and the Coulomb effects, which give important properties of the calculated SD bands in nuclei around $^{32}$S. Then, a classification of the SD bands and a description of the band crossings is presented.

## A. Deformed shell gaps

The ground state of the $^{32}$S nucleus, obtained within the HF approach with the SLy4 force, corresponds to a spherical-shape configuration that contains, on top of the closed $^{16}$O core, the $1d_{5/2}$ and $2s_{1/2}$ orbitals filled, both for the neutrons and protons. With an increase in the prolate deformation, the negative-parity Nilsson orbitals originating from the spherical $^{16}$O core stay occupied (see Fig. 2.21a in Ref. [39] for a qualitative illustration). The same is true for the positive-parity valence orbitals, except for the orbital [202]5/2 (the up-sloping extruder orbital), which originates from the spherical $1d_{5/2}$ shell, and rapidly grows up in energy with increasing deformation. After this orbital is crossed by the [330]1/2 orbital (the down-sloping intruder orbital), which originates from the spherical $1f_{7/2}$ shell, one obtains a large, about 2.5 MeV gap that corresponds to a SD configuration in the $^{32}$S nucleus. Therefore, the SD states in such a light system as $^{32}$S, formally correspond to the 4p-4h excitation with respect to the spherical ground state. However, our calculations presented in detail below indicate that the SD configurations have extremely large quadrupole deformations, $\beta \simeq 0.7$, and the implied structures of the SD, ND (normal deformed), or spherical wave functions have so little similarities that the notion of particle-hole excitations with respect to the spherical ground-state is not very useful.

Figures 1 and 2 show the neutron and proton single-particle routhians, as functions of the cranking frequency $\hbar\omega$. One can see that over a very large range of the rotational frequencies, there exists an important gap in the single-particle HF spectrum at the neutron and proton numbers $N=Z=16$. By definition, in the underlying $^{32}$S SD configuration all the neutron and proton levels lying below the gaps at $N=16$ and $Z=16$ are occupied, and all those above the gaps are empty.

As a result of the presence of those large gaps in the single-particle $^{32}$S proton and neutron spectra, we refer to the corresponding lowest-energy SD state as to the magic SD configuration.

A characteristic result visible from Figs. 1 and 2 is that the over-all single-particle structure of the HF orbitals near the Fermi level is remarkably simple. First of all, the dependence of the single-particle routhians on the rotational frequency is very regular, and there is only one clear-cut crossing caused by the down-sloping routhians $[440]1/2(r=-i)$, originating from the $N_0=4$ shell. Second, the density of levels appearing in the figures is very low as compared, e.g., to those in the mass $A \simeq 150$ region of SD nuclei. The negative parity states are represented only by two $N_0=3$ intruder orbitals $[330]1/2(r=\pm i)$ below, and two intruder orbitals $[321]3/2(r=\pm i)$ above the Fermi level. Similarly, in the positive parity there are only two states $[211]1/2(r=\pm i)$ below, and two extruder states $[202]5/2(r=\pm i)$ above the Fermi level. Signature splitting of the extruder states $[202]5/2(r=\pm i)$ is very weak, because they carry high $K=5/2$ angular momentum projection, whereas splitting between the intruder levels $[321]3/2(r=\pm i)$ is more pronounced. It becomes well visible at rotational frequencies of about 0.8 MeV. Below the Fermi level, orbitals $[330]1/2(r=\pm i)$ and $[211]1/2(r=\pm i)$ have $K=1/2$, hence both are strongly split.

## B. Coulomb effects and isospin symmetry

An important observation that follows from a comparison of Figs. 1 and 2 is that the neutron and proton routhian spectra are almost identical, apart from a nearly constant shift in energy that amounts to about 6 MeV. Such a constant shift expresses the fact that despite a possibly non-trivial evolution of the individual-nucleonic wave functions in terms of the rotational frequency, the corresponding Coulomb interactions average out to nearly a constant, and correspond to the Coulomb energy of a rotating but otherwise $\hbar\omega$-independent charge distribution.

On the one hand, degeneracy of neutron and proton routhians is a manifestation of a charge-independence of the forces used. On the other hand, however, the pairs of nearly degenerate proton and neutron wave-functions may be used to introduce an alternative representation in terms of the isoscalar- and isovector-coupled wave-functions. In such a case any arbitrary isospin-symmetric residual interaction is likely to introduce systematic differences in the spectra of the $T=0$ and $T=1$ states. This would allow to test that particular component of the forces against experiment - or, conversely, from an existence of systematic discrepancies between experiment and mean-field calculations - it would allow to optimize the residual interactions. The observed near-degeneracy of the corresponding proton and neutron levels is in fact a prerequisite indication that in this mass region the isospin-symmetry effects could be very important. We will use the above observation as a guideline in further analysis of the neutron/proton configurations in $^{32}$S and neighboring nuclei.

## C. Classification of SD bands

For the conserved parity, $\pi=\pm$, and signature, $r=\pm i$, quantum numbers, the space of single particle states is separated into four parity/signature blocks, $[\pi,-ir]=(++,+-,-+,--)$. By constructing a particle-hole excitation we necessarily arrive at a rearrangement among the four blocks of levels - one class of rearrangements always leading to the occupation of all the lowest levels *within* each block. It turns out that such states form a majority among the low-lying bands studied here. Supposing that the lowest states in each of the blocks are occupied, one may describe in the standard way the many-particle configurations by giving the numbers of states occupied in each block. In this notation, the SD $^{32}$S magic configuration is given by the (4,4,4,4) occupation numbers, both for neutrons and protons, while the ground-state configuration reads (5,5,3,3).

All configurations that are examined below are built by exciting particles from the four levels below, to the four levels above the neutron and/or proton Fermi energies at the SD shape. The remaining orbitals below the Fermi levels will always be occupied. Therefore, the single-particle neutron or proton active spaces are composed of 8 orbitals (4 intruders and 4 non-intruders) that contain 4 particles. This leads to $C_4^8 = 70$ possible many-body SD configurations for neutrons and $C_4^8 = 70$ SD configurations for protons. The fact that among the bands studied in this article always the lowest states in each parity/signature block are occupied (other cases, where necessary, will be explicitly mentioned) reduces these numbers from 70 to 19 neutron or proton configurations necessary to control the low-energy rotational bands constructed within the discussed active spaces.

Further, we use the observation that for both intruder states, [330]1/2 and [321]3/2, the $r=+i$ signature partners are always below the $r=-i$ partners (for all deformations and rotational frequencies). Hence, the intruder orbitals should preferably be occupied in that order of increasing energy, i.e., when one particle occupies the negative-parity orbitals, it will occupy the [330]1/2($r=+i$) orbital, when two – they will occupy the [330]1/2($r=+i$) and [330]1/2($r=-i$) orbitals, and when three - they will occupy the [330]1/2($r=+i$), [330]1/2($r=-i$), and [321]3/2($r=+i$) orbitals, etc. This rule reduces the number of available configurations from 19 to 9. Finally, we reject two more configurations, as described below, and we are left with 7 configurations to be considered for neutrons and for protons. Although such a preselection of configurations may appear to be quite arbitrary, it is in fact based on the requirement that one wants to end up with a restricted set of low-energy configurations only.

Figure 3 shows schematically the single-particle orbitals (top), as well as all the considered here particle-hole configurations (bottom). The same diagram is valid both for neutrons and protons. The four intruder-states that are close to the Fermi energies are all characterized by the principal HO quantum number $N_0$=3. Following the well-established notation, we denote the neutron or proton intruder occupations by the symbol $3^{n/p}$, where $n$ or $p$ are the numbers of the occupied neutron and proton intruder states, respectively.

As illustrated in the figure, for the $3^0$ or $3^4$ configurations, there are four or none particles, respectively, in the positive-parity states, and hence, in our predefined phase space, these configurations are unique. For the $3^2$ configuration, two particles occupy the positive-parity states, and we restrict our considerations to only one (out of three) occupation variant, namely, we require both particles to occupy the two signatures of the lower orbital [211]1/2. Hence, in the following, symbol $3^2$ pertains to this particular configuration. Finally, if one or three intruder states are occupied, i.e., in the $3^1$ or $3^3$ configurations, there are accordingly, three particles or one particle in the positive-parity orbitals, and here an additional label is necessary. We distinguish the corresponding configurations by introducing subscripts + or −, i.e., by using symbols $3^1_+$, $3^1_-$, $3^3_+$, $3^3_-$. The subscripts correspond, (i) in the $3^3$ configurations, to the signature of the occupied [211]1/2($r=\pm i$) orbital, and (i) in the $3^1$ configurations, to the signature of the occupied [202]5/2($r=\pm i$) orbital. Whenever symbols $3^1$ or $3^3$ without the subscripts are used, they pertain to both such configurations.

After having preselected the 7 neutron and proton configurations, we have at our disposal 49 configurations of the whole nucleus, which we denote by $3^n3^p$, and when necessary supplement by the signature subscripts + or −, as described above. For example, the magic SD configuration of $^{32}$S is denoted by $3^23^2$, and the ground-state configuration reads $3^03^0$.

A manifest symmetry between neutrons and protons implies a manifest symmetry between the corresponding rotational bands. We have verified that those bands which are mirror images obtained from one another by replacing the neutrons by the protons and *vice versa* lead to almost identical results. However, because of the larger spatial extensions of intruder orbitals as compared to positive-parity orbitals, the (very small) Coulomb shifts will always slightly bring down the $p>n$ configurations below those with $p<n$. Consequently, in the following we consider only the $3^n3^p$ configurations for $p \geq n$. Introducing these last arguments into our selection scheme, we end up with 30 $^{32}$S configurations to be considered in the further analysis.

## D. Level crossings and the HF convergence

In all calculations, we diabatically follow configurations, i.e., we always occupy orbitals with given single-particle characteristics, irrespectively of whether they cross with the other orbitals or not. This is technically easy if crossings involve orbitals of different parity/signature blocks, but special techniques [36] must be used in self-consistent methods, to diabatically follow configurations which cross within the common parity/signature blocks. Those crossings are particularly interesting because they usually give rise to an up-bending or back-bending structures, and are thus important for the experimental identification of the underlying structures.

In the present study we separate the diabatic configurations by proceeding as follows. If, in the positive-parity orbitals, a particle "switches on and off" from the occupation of one orbital to another, we force an occupation of the state that has a *larger* single-particle alignment, independently of whether it is slightly higher or lower in energy. In the present case this implies that we always force the particle into the [211]1/2 orbital and leave the extruder orbital [202]5/2 empty. Incidentally, by occupying the state that has a *smaller* alignment, at the end of a successful iteration sequence we obtain a markedly different solution, with much smaller deformations, i.e., the fact that two configurations mix does not imply that both manifest all the same physical properties in this case.

A rich collection of the experimental data on the back-bending and up-bending phenomena that exist today in the literature has been interpreted in terms of the theoretical single-particle configurations followed according to the diabatic scheme. Whether the experimental bands exist that follow an adiabatic scheme is an open question, and an unambiguous (while anyway model depended) demonstrations are very difficult. On the theory grounds, this question cannot be settled within a mean-field approach. Therefore, our approach to follow diabatic configurations is dictated by the fact that a great majority of the high spin data has been interpreted accordingly. In case of need demonstrated by future experiments, our present results could immediately be used as a first step in the band-band mixing calculations.

By examining the routhian diagrams obtained self-consistently at a fixed particle-hole configuration, as, e.g., those shown in Figs. 1 and 2, one cannot predict crossings which may happen in some other configuration. This is especially true in $^{32}$S, where different configurations correspond to fairly different deformations, see Sec. IV B, and therefore, may involve significantly different ordering of orbitals. As an example, in Figs. 4 and 5 we present neutron routhians corresponding to the self-consistently calculated $3^1_- 3^1_-$ and $3^1_+ 3^1_+$ bands, respectively. Since these configurations produce deformations significantly smaller than that of the magic SD configuration, the (empty) extruder orbitals [202]5/2($r=\pm i$) are here much lower in energy, and strongly mix with the (occupied) [211]1/2($r=\pm i$) orbitals. In these configurations, the signature splitting is very large, and strongly depends on the actual configuration, see discussion in Sec. IV D. Therefore the order of routhians in the $3^1_+ 3^1_+$ configuration (Fig. 5) is entirely different from that in the $3^1_- 3^1_-$ configuration (Fig. 4) and leads to very strong mixing and level repulsion at and near the crossing frequency in the latter case.

We can easily identify the crossing regions by the simple fact that the HF iterations are poorly convergent, or non-convergent there [36]. This concerns only those methods of solving the HF equations, which are based on successive diagonalizations of the mean-field Hamiltonian. The gradient method and the imaginary-time method [39], always arrive at the smallest-energy solution, and do converge. However, the obtained solutions simply correspond to one of the infinitely many possible mixed-orbital solutions, with the same or very close energy, and consequently, those methods do not cure the problem, but allow for not seeing it.

We have made every possible effort to achieve convergence of all configurations at all angular frequencies, however, in several cases it turned out not to be possible. This is the case, for example, for the $3^1_+3^1_+$ band at $\hbar\omega$=1.0–1.4 MeV; the non-convergence here results in an irregular behavior of routhians in Fig. 2. In the following, we show energies corresponding to the non-converged points along with the well-converged ones, however, we remove points corresponding to non-converged solutions from plots of other observables.

## IV. SUPERDEFORMED BANDS IN $^{32}$S

In this section we present results for the energies of the rotational bands, and discuss other effects and observables, i.e., shape evolution with spin and shape coexistence, signature-related degeneracies, dynamical moments, and relative alignments.

### A. Energies

In Fig. 6 are plotted the HF energies as functions of spin for the 30 SD bands calculated in $^{32}$S. As it is often done in the cranking approach, we identify the average projection of the angular momentum on the cranking axis $\langle I_y\rangle$ with the total angular momentum of the system, i.e., we set $I=\langle I_y\rangle$. (Within a more refined approximation some authors identify $I(I+1)$ with $\langle I_y\rangle^2$ [39], what results in a standard (approximate) correction $I=\langle I_y\rangle-\frac{1}{2}\hbar$; however, this is not implemented in the figures presented below.)

All the bands have been obtained within the cranking HF formalism, with the rotational frequencies that start at $\hbar\omega$=0.4 MeV and increase in steps of 0.2 MeV. For each band, the calculations were carried out up to the highest rotational frequencies that did not induce any sudden configuration change. Since almost all bands are crossed at high rotational frequencies by the bands involving the down-sloping $[440]1/2(r=-i)$ orbital, cf. Figs. 1 and 2, and Ref. [28], such configuration changes are in many cases inevitable. On the one hand, introducing an upper limit of the frequencies of some calculated bands reflects a deficiency of the method since the discussed crossings are in general the physical ones. On the other hand, however, the corresponding experimental results are expected to deviate from regularity at the vicinity of the calculated limiting $\hbar\omega$ values and are likely to manifest, e.g., an up- or even a back-bending behavior there, thus offering a possibility of valuable tests of the crossing frequencies anyway.

Let us remark in passing that within the HO model, when two protons and two neutrons in the HO $[440]1/2$ states are added to the $^{32}$S SD configuration, one obtains the magic hyperdeformed HO configuration in $^{36}$Ar. Structures based on the $[440]1/2(r=-i)$ orbitals are abundant in $^{32}$S, however, they should rather be attributed to the hyperdeformed configurations, and are not studied in the present article.

Bands shown in Fig. 6 have been separated into four groups, plotted in four panels. Figures 6(a), (b), (c), and (d) show the $3^n 3^p$ bands with $p=n$, $p=n+1$, $p=n+2$, and $p \geq n+3$, respectively. Parities of bands are equal to the products of parities of the proton and neutron configurations, i.e., $\pi = (-1)^{n+p}$ in our case, and are denoted by full ($\pi=+1$) and open ($\pi=-1$) symbols. Various forms of the symbols (circles, squares, etc.) distinguish different values of $n$. In order to further differentiate between various configurations, we have to introduce a convention relating the line styles with the signatures of neutron *and* proton subsystems, $r_n$ and $r_p$. Hence, long-short-dashed, solid, dotted, and dashed lines denote $(r_n,r_p)=(+,+)$, $(+,-)$, $(-,+)$, and $(-,-)$ signatures, respectively. Of course the total signature $r$ of each band is always equal to $r=r_n \times r_p$.

Since we are mostly interested in the low energy configurations, in Fig. 7 we show a blow-up of the near-yrast region of energies, for a selection of bands being closest to the yrast band. At $I = \langle I_y \rangle - \frac{1}{2}\hbar = 6\hbar$ (with the standard spin correction of $\frac{1}{2}\hbar$ here subtracted), we obtain in the $3^0 3^0$ band the total energy of $E=-261.651$ MeV, which gives the calculated excitation energy of $E_x(I=6\hbar)=8.349$ MeV, ridiculously close to the experimental energy, 8.346 MeV [31,32], of the $6^+$ yrast state in $^{32}$S. Of course, an agreement on this level of precision is to a large extent accidental, however, it gives us confidence that a correct configuration is being followed at low excitation energies.

At low spins, the yrast line is first built upon the ground-state $3^0 3^0$ configuration whose energy increases very regularly up to the angular momentum of $I \simeq 9\hbar$ and excitation energy of $E_x \simeq 14.5$ MeV. At $I \simeq 9$–$10\hbar$, the one-intruder configurations $3^0 3^1$ become yrast for a narrow region of spins. These bands are next crossed at $E_x \simeq 16.2$ MeV by two bands with $r=+1$, the $3^1 3^1$ configurations, which are yrast up to about $I \simeq 15\hbar$. At this point ($E_x \simeq 25$ MeV) the yrast line has the structure of the magic $3^2 3^2$ SD configuration. When extrapolated to zero spin, the magic SD configuration corresponds to the excitation energy of $E_x \simeq 14$ MeV.

The spin (energy) range of up to $I \sim 6\hbar$ ($\sim 10$ MeV) can be very well described by the *sd* shell-model calculations [32], and in the rest of this article we will focus on the higher spin states. [Some collective bands in the low spin (energy) range may be unstable with respect to parity-breaking deformations [28]; we do not study those effects either.]

### B. Quadrupole moments

Using the same symbols as those introduced in Fig. 6, in Fig. 8 are plotted the proton quadrupole moments, in the form of trajectories of points on the $Q_{20} - Q_{22}$ plane, corresponding to consecutive values of the rotational frequency. In order to visualize the fact that values of $Q_{22}$ are always much smaller than those of $Q_{20}$ (small non-axiality), the lines corresponding to $\gamma = \pm 15°$ and $\gamma = \pm 30°$, where $\tan(\gamma) = Q_{22}/Q_{20}$, are also shown in the figure.

From Fig. 8 it is clearly seen that bands calculated in the present study represent fairly distinct regions of deformation. In order to better visualize the magnitude of the deformation, one can use the simplest first-order formula [39], $\beta = \sqrt{5/\pi} Q_{20}/(Ze\langle r^2 \rangle)$, relating the proton axial quadrupole moment with the standard deformation parameter $\beta$. For the $3^2 3^2$ configuration this gives $\beta \simeq Q_{20}/(2.53 \, \text{eb}) \simeq 0.7$. Since at the same time the axial hexadecapole moment is fairly small, $Q_{40} \simeq 0.06 \, \text{eb}^2$, the first-order formula should be a good estimate of the exact result, corresponding to the deformations of an equivalent sharp-edge uniform charge distribution that has all multipole moments equal to the ones calculated microscopically.

It follows that the ground-state band $3^03^0$ reaches quadrupole deformations $\beta$ of the order of 0.16, the intermediate-deformation configurations $3^13^1$ correspond to $\beta\simeq0.45$, while the strongest deformed band $3^43^4$ has $\beta\simeq0.8$. (The latter band carries, however, $Q_{40}\simeq0.54\,\text{eb}^2$, and thus the simple one-parameter formula for $\beta$ can be less precise here.) Results presented in Fig. 8 show a clear correlation of quadrupole moments and numbers of intruder states occupied in a given configuration; we discuss this question in more detail in Sec. V.

### C. Signature-splitting

In Figs. 6 and 7, there are several pairs of nearly degenerate bands. First of all, all the $p=n+1$ and $p=n+3$ bands shown in Figs. 6(b) and (d), form usual pairs of signature-partner bands differing by signatures of odd nucleons. The signature splitting of these partner-bands closely follows the signature splitting of the corresponding positive-parity single-particle routhians, see Figs. 1 and 2. Indeed, in bands $3^03^1$ or $3^13^2$, for example, the odd neutrons or protons, respectively, occupy orbitals $[202]5/2(r=\pm i)$. These orbitals show almost no signature splitting, and hence almost no signature splitting is also seen in Fig. 6(b) (circles and squares). Similarly, in bands $3^23^3$ and $3^33^4$, the signature-split $[211]1/2(r=\pm i)$ orbitals are occupied, and this gives similarly signature-split pairs of bands (diamonds and triangles). The same pattern is repeated for the $3^03^3$ and $3^13^4$ pairs of bands in Fig. 6(d). [Incidentally, not always both signature partners can be followed up to the same spin; for example, the $3^03^1_-$ band continues to a higher spin than its partner band $3^03^1_+$, because for the former band, the deformation significantly changes at high rotational frequencies, see Fig. 8(b).]

### D. Signature-separation sensitive to the time-odd channels

A different situation takes place in configurations where both a neutron *and* a proton occupy unbalanced-signature states. In particular, four near-yrast $3^13^1$ configurations, shown in Figs. 6(a) and 7, group into two nearly degenerate pairs of bands having the same signature. Indeed, the $r=+1$ bands, $3^1_-3^1_-$ and $3^1_+3^1_+$, are very close to one another, with the latter one lying slightly lower in energy, in accordance with the sign of the small signature splitting of the high-$K$ $[202]5/2(r=\pm i)$ orbitals. Note that the sudden deviation from regular behavior, seen in the latter band at $\hbar\omega=1.0$–$1.4$ MeV, is due to a poor HF convergence related to strong mixing of orbitals, see discussion in Sec. III D.

The second pair of degenerate $3^13^1$ bands corresponds to the signature $r=-1$, and is composed of the $3^1_-3^1_+$ and $3^1_+3^1_-$ configurations, that are the mirror partners of one another in terms of the isospin. Therefore, irrespectively of the small signature splitting of the $[202]5/2(r=\pm i)$ orbitals, these bands are almost perfectly degenerate. Again, due to the interactions between orbitals, at $I\simeq10$ and $I\simeq14$ one observes small irregularities reflecting poor HF convergence.

A remarkable feature obtained in the HF calculations is the fact that the pair of bands just mentioned, with $r=-1$, lies about 2 MeV above the $r=+1$ pair. As opposed to the standard signature slitting effect, we my call these bands the signature-separated bands. Such a separation could not have been obtained in a phenomenological cranking model, i.e., the one using the Woods-Saxon or Nilsson potentials, because there the single-particle degeneracies immediately imply degeneracies of bands in many-body systems. Indeed, by putting one neutron and one proton into weakly split and non-interacting [202]5/2($r=\pm i$) pair of orbitals, one should have obtained all the four $3^1 3^1$ bands strongly degenerate. Note that the separation of the $r=-1$ and $r=+1$ pairs of bands cannot be due to the deformation effect, because deformations of the four bands are very similar, see Fig. 8(a).

Strong separation of pairs of signature-degenerate bands results [40,41] from the self-consistent effects related to the time-odd components [34] in the HF mean fields. Odd particles induce the time-odd mean fields in odd and odd-odd nuclei, and similarly, odd particles in signature-unbalanced states induce the time-odd mean fields for certain configurations of even-even nuclei. In particular, when a neutron is put into the [202]5/2($r=+i$) orbital, it creates, through the strong neutron-proton interaction which is inherent to any effective nuclear force, e.g., to the Skyrme force, a strong attractive component in the proton mean-field corresponding to the same, i.e. $r=+i$ symmetry. Therefore, when the proton occupying the [202]5/2($r=+i$) orbital is put into such a mean field, the total energy is significantly lowered. Of course, exactly the same mechanism applies for two particles occupying the [202]5/2($r=-i$) orbitals. The proton mean field, generated by the [202]5/2($r=+i$) neutron, does not influence the states of the $r=-i$ symmetry, and therefore, adding then the [202]5/2($r=-i$) proton does not influence the total energy. Hence, here the $r=-1$ bands are not affected by the time-odd interactions, (i.e., the interactions which give the time-odd mean fields through the HF averaging), while the $r=+1$ bands are significantly affected, and acquire an additional binding.

Obviously, the magnitude of the separation between the $r=-1$ and $r=+1$ bands crucially depends on the interaction strengths in time-odd channels. Unfortunately, the coupling constants corresponding to these channels [34] are not restricted by typical ground-state observables (masses, radii, etc.), that serve as experimental benchmarks with respect to which the force parameters are adjusted. Therefore, high spin effects, like the aforementioned $r=-1$ vs. $r=+1$ separation, that are manifestly sensitive to these unexplored channels of the interaction, could provide an extremely important information pertaining to basic properties of nuclear effective forces. Note that in $^{32}$S, the structure of the yrast line dramatically depends on the strength of the interaction in these channels, because the $3^1 3^1$ bands become yrast mostly due to the time-odd interaction.

By looking at similar quartets of bands, e.g., those corresponding to the $3^3 3^3$ and $3^1 3^3$ configurations, Figs. 6(a) and (c), respectively, one sees that the strength of interactions in the time-odd channels depends on the structure of the underlying orbitals. The signature splitting of the [211]1/2($r=\pm i$) orbitals obscures the picture a little because it gives the splitting of the $3^3_+ 3^3_+$ and $3^3_- 3^3_-$ configurations, however, the centroid of these two configurations lies visibly below the perfectly degenerate pair of mirror partners $3^3_- 3^3_+$ and $3^3_+ 3^3_-$. Hence, within the [211]1/2($r=\pm i$) orbitals, the time-odd interaction is significantly weaker. Finally, there seems to be no such a non-diagonal interaction between the [211]1/2($r=\pm i$) and [202]5/2($r=\pm i$) orbitals, because the degeneracy pattern of the $3^1 3^3$ orbitals is completely different. Indeed, the standard, weakly split, two signature pairs appear, the lower one composed of the $3^1_+ 3^3_+$ and $3^1_- 3^3_+$ configurations, and the higher one composed of the $3^1_+3^3_-$ and $3^1_-3^3_-$ configurations, in accordance with the sign of the signature splitting of the [211]1/2($r=\pm i$) orbitals.

Results presented in this section indicate that the properly selected high-spin structures in a SD nucleus reflect the properties of the effective interaction in the time-odd channel. Quantitatively, in the restricted set of orbitals considered in $^{32}$S, the time-odd interaction amounts to an attractive force which acts between protons and neutrons occupying *the same* orbitals, i.e., orbitals having the same quantum numbers. Therefore, the discussed interaction channel has several features of the $T=0$ pairing interaction, although obviously the effects discussed here are not related to any collective pairing channels, but rather pertain to interactions in the particle-hole channel.

### E. Dynamical moments and relative alignments

A mixing of two common-symmetry orbitals that approach each other at the Fermi energy creates non-converged HF solutions for certain values of $\hbar\omega$ as discussed in Sec. III D, and introduces large errors in the observables calculated in this article except, perhaps, for energies and multipole moments. Indeed, in many cases of non-converging solutions, due to the variational character of the HF equations, the total energies are almost correct, namely, they can be smoothly followed through the crossing region. However, errors in the total spins can be much larger, because the non-converged solutions correspond to almost-random mixtures of two interacting orbitals that are very close in energy but may significantly differ in spin. Then, neither the relative alignments, nor, especially, the dynamical moments, can be smoothly followed along the crossing region. Therefore, in the figures presented in this section, we removed all points corresponding to the non-converged solutions; the absence of some points was compensated for by drawing straight lines between points corresponding to the converged solutions.

In Fig. 9 are reported the dynamical moments $\mathcal{J}^{(2)}=dI/d\omega$, calculated for the near-yrast bands in $^{32}$S. One can see, that bands with the same intruder contents present very similar behavior, as far as the dynamical moments are concerned. It appears clearly from the figure that bands based on the $3^1_\pm$ and/or $3^0$ configurations have in general (especially at high rotational frequencies) much lower dynamical moments than the magic $3^23^2$ SD band. The bands based on higher numbers of occupied intruder states (not shown in the figure), have higher values of $\mathcal{J}^{(2)}$, along with a larger distance from the yrast line.

In Fig. 10 are drawn the relative alignments, $\delta I=I(\text{band})-I(\text{SD }3^23^2\text{ band})$, of near-yrast bands in $^{32}$S calculated with respect to the magic $3^23^2$ band in the same nucleus. One can see that again the results obtained for various bands depend mainly on the numbers of occupied intruder states. It is very difficult for the nucleus to build up spin, when few intruder orbitals are occupied, and therefore one observes a lowering of the relative alignments for these bands. One may discuss these questions more clearly by introducing the relative alignments of bands in neighboring nuclei, presented in Sec. V.

All the calculated features of SD bands in $^{32}$S seem to reflect in a very direct way the crucial role played by the intruder orbitals. Such an observation may, therefore, similarly as in other SD regions, serve as a guideline in theoretical analyses, as well as in experimental investigations of properties of SD bands in the $A\simeq30$ mass region.

# V. SUPERDEFORMED BANDS IN ONE-PARTICLE AND ONE-HOLE NEIGHBORS OF $^{32}$S

In order to analyze polarization effects induced by individual particle or hole orbitals in $^{32}$S, we have also performed the HF calculations for the four neighboring nuclei: $^{33}$S, $^{31}$S, $^{33}$Cl, and $^{31}$P. Among them, there are two pairs of mirror nuclei, $^{33}$S – $^{33}$Cl and $^{31}$S – $^{31}$P. For each of these nuclei we have calculated four bands, corresponding to either the four lowest available particle states (in $A$=33), or the four highest available hole states (in $A$=31). In other words, in $^{33}$S or $^{33}$Cl the neutron or proton is added to the magic $3^2 3^2$ $^{32}$S configuration, in the $[321]3/2(r=\pm i)$ and $[202]5/2(r=\pm i)$ orbitals, which gives the neutron or proton configurations: $3^2_+$, $3^2_-$, $3^3$, and $3^{3*}$. Here, by an asterisk we denote the configuration in which a particle is added not to the lowest available intruder state, but to the next-to-lowest available intruder state. Similarly, in $^{31}$S and $^{31}$P the neutron or proton is removed from the magic $3^2 3^2$ $^{32}$S configuration, from the $[330]1/2(r=\pm i)$ and $[211]1/2(r=\pm i)$ orbitals, which gives the neutron or proton configurations: $3^2_+$, $3^2_-$, $3^1$, and $3^{1*}$. Again, by an asterisk we denote the configuration in which a particle remains not in the lowest available intruder state, but in the next-to-lowest available intruder state.

In Figs. 11 and 12 we show energies of the calculated HF bands in $^{33}$S, $^{31}$S, $^{33}$Cl, and $^{31}$P. One can observe that the mirror nuclei have extremely similar SD spectra. Bands in the $A$=33 nuclei form pairs of degenerate signature partners, while those corresponding to the signature partners in $A$=31 are strongly split, in accordance with the characteristic features of the corresponding single-particle routhians, Figs. 1 and 2. Note that the ground state bands in the $A$=33 nuclei correspond to non-intruder particle states, and similarly, those in the $A$=31 nuclei correspond to holes in non-intruder orbitals.

The HF calculations give the energies of rotational bands on the absolute scale. Therefore, in order to estimate the available $Q$-value windows for particle emissions, one may simply compare (at a given value of the angular momentum) the energies shown in Figs. 6, 11, and 12. Since the rigid-rotor reference energies are the same at fixed spins, one can directly compare the values given in the figures. For example, for $^{31}$S the yrast energy at $I$=12 is about $-244$ MeV, which shows that none of the $^{32}$S bands shown in Fig. 6, except the $3^4 3^4$ and $3^0 3^4$ configurations, can emit a zero-angular-momentum neutron to the SD states in $^{31}$S. Similarly, for $^{31}$P the corresponding yrast energy is $-249.5$ MeV, which opens up the proton emission channel from several other bands, but not those from the near-yrast bands shown in Fig. 7.

Let us emphasize that the angular momentum, $\ell$, carried away by an emitted particle, dramatically influences the considered $Q$-values in nuclei around $^{32}$S, especially at high spins. Since after subtracting the rigid-rotor reference, the energies of bands are fairly flat (Figs. 6, 11, and 12), one can very simply estimate the $Q$-values at given $I$ and $\ell$ values to be by an amount of $[2I\ell+\ell(\ell+1)]\times 0.05$ MeV larger than those at $\ell$=0. For instance, at $I$=20, and with the angular-momentum transfer $\ell$=2 (or 3), the additional energies in a daughter nucleus are 4.3 (or 6.6) MeV. Consequently, the protons emitted through the high angular-momentum (e.g., $N_0$=3) orbitals are among the most likely candidates for the band-to-band emission mechanism. From the results presented in the figures one may precisely estimate the $Q$-value windows for particles carrying out any given amount of the angular momentum from any given band.

The illustrations of the dynamical moments in $^{33}$S, $^{31}$S, $^{33}$Cl, and $^{31}$P, shown in Figs. 13 and 14, indicate an extreme similarity of the results in mirror nuclei. This suggests that several among the SD bands in the mirror nuclei around $^{32}$S might manifest the "identical band" phenomenon. Comparing these results with those in the magic SD band in $^{32}$S, Fig. 9, one sees that particles in the intruder [321]3/2 orbitals and extruder [202]5/2 orbitals, respectively add and subtract $1\hbar^2$/MeV(at high spin) with respect to the magic core. Variations of $\mathcal{J}^{(2)}$, that correspond to the intruder and non-intruder hole states, are of the similar order.

By calculating differences between one-body observables, like the angular momentum or quadrupole moment, determined in $^{33}$S, $^{31}$S, $^{33}$Cl, and $^{31}$P, and in $^{32}$S, one can identify basic single-particle properties of all important orbitals around the SD $^{32}$S magic-core configuration. These differences correspond not only to the bare average values of the observables, calculated for given orbitals, but also include complete polarization effects. It is known that in the SD $A \simeq 150$ nuclei, the single-particle alignments [42], and charge quadrupole moments [43,44], constitute additive quantities with respect to adding and subtracting particles from the magic SD configurations of $^{152}$Dy. An analogous observation is also confirmed by calculations in the SD $A \simeq 60$ nuclei [29,45,46]. In the present paper we have verified the additivity of alignments and quadrupole moments between the SD bands in $^{32}$S, and in $^{33}$S, $^{31}$S, $^{33}$Cl, and $^{31}$P. Tests of this principle in other nuclei around $^{32}$S are left for a future publication.

In Figs. 15 and 16 we present the obtained relative alignments $\delta I$ and proton quadrupole moments $\delta Q_{20}$, respectively. Since the relative alignments pertain to the total angular momentum, the effects of neutron and proton orbitals, obtained in the $N=16$ and $Z=16$ nuclei, respectively, are almost identical. For relative proton quadrupole moments, the effects of neutrons and protons are different, because neutrons contribute only through the polarization effects, while for protons one also has the bare direct contribution. In Figs. 15 and 16 we also indicate by which particle- or hole-orbital differ the bands in $^{33}$S, $^{31}$S, $^{33}$Cl, and $^{31}$P form the magic SD band in $^{32}$S.

One can see that the relative alignments generated by various orbitals differ considerably. Therefore, the relative alignments may serve as distinct fingerprints of orbitals in SD nuclei around $^{32}$S. In particular, the second intruder, hole-orbital $[330]1/2(r=-i)$, gives rather large negative relative alignment, while the positive-parity, hole-orbital $[211]1/2(r=+i)$, gives a rather constant alignment of about $-1\hbar$, and hence may be at the origin of hypothetical yet another class of identical bands in this region.

The relative proton quadrupole moments of orbitals around the magic $N=Z=16$ SD gap are fairly constant in function of the rotational frequency. Values corresponding to intruder orbitals are usually much larger than those corresponding to positive-parity orbitals. Hence, one can easily understand the origin of groups of $^{32}$S bands having significantly different quadrupole moments, see Sec. IV B. As far as the polarization effects alone are concerned, the extruder particle orbitals $[202]5/2(r=\pm i)$ carry almost the same effect as the intruder hole-orbitals $[330]1/2(r=\pm i)$. Needless to say, these are the main orbitals which are at the origin of the SD shapes in $^{32}$S.

# VI. STABILITY OF THE SD CONFIGURATIONS AROUND $^{32}$S

It is often possible to discuss the stability of the SD configurations with the help of the total energy surfaces obtained with the Strutinsky or constrained HF methods. According to such a representation, high barriers surrounding a potential minimum are usually interpreted as a sign of a large stability of a given nucleus against, e.g., fission or shape transitions.

Strictly speaking, the physical solutions obtained with the HF method correspond to a discrete set of local minima of the HF functional. Using the language of the simple deformed shell-model: the HF minima obtained in $^{32}$S nuclei are strongly separated in terms quadrupole moment treated as a measure of the deformation. By using the constrained HF approach we could in principle always connect those isolated points thus obtaining potential barriers analogous to the ones obtained within the Strutinsky method. However, the physical interpretation of the results should be different depending on whether very many or only very few intermediate configurations are available for a given physical system. When many solutions are densely distributed along the deformation axis, the physical system is likely to undergo a sequence of transitions between the states that differ in deformation only a little, and the Strutinsky as well as HF results can be interpreted as physically analogous. Such a situation takes place, e.g., in the SD nuclei in the $A\simeq 150$ and $A\simeq 190$ mass regions.

In nuclei from the vicinity of $^{32}$S, the occupying or not occupying just two intruder orbitals makes a significant difference in terms of the quadrupole moments of the resulting HF solutions. As a consequence the potential energy surface (PES) representation (see Fig. 8 of Ref. [26] for the PES in $^{32}$S) is most likely not the best way of getting the information about the stability of the SD configurations with respect to a decay into any other shape configuration. Indeed, the decay will be in general hindered by a difference in configurations between the initial and the final states. Such a difference remains totally invisible from, e.g., the $E$ vs. $Q_{20}$ sequence of constrained HF (or HFB) solutions, which all correspond to a different mixing of merely two configurations.

The above remarks apply independently of the following, more general observation: the barrier pictures may become often strongly misleading because the barrier extensions (shapes) do not carry any direct physical relation to the behavior of the object studied. A useful physical meaning can be attributed to those objects only *after* having introduced a description of the collective inertia adapted to the deformation space in use. Such a description, either obtained within the generator coordinate method, or described in terms of the collective inertia tensor, implicitly takes into account the slowing down of the transition caused by the aforementioned configuration changes.

# VII. CONCLUSIONS

In this article, consequences of the predicted existence of the $N=16$ and $Z=16$ strong superdeformed shell closures in the $^{32}$S nucleus, together with the role of the close-lying intruder orbitals, are analyzed and discussed.

The calculated proton and neutron single-particle spectra in $^{32}$S turn out to be nearly identical, apart from an approximately constant shift of about 6 MeV. As a consequence, several rotational bands in nuclei around $^{32}$S are predicted to produce an "identical band" effect, and the corresponding results are discussed in some detail.

The property of additivity expressed, e.g., in terms of multipole moments, that was discovered originally in heavier SD nuclei, is confirmed to hold also for the $^{31,32,33}$S, $^{31}$P, and $^{33}$Cl nuclei. In these five nuclei, detailed predictions related to the dynamical moments and relative alignments are also illustrated. Similarities and differences between properties of various bands are discussed and criteria facilitating an identification of some characteristic excited configurations and single-particle orbitals are formulated.

It is pointed out that the time-reversal symmetry-breaking in the self-consistent HF mean-field can manifest itself through a strong separation between the bands that in a standard Nilsson approach must appear as nearly degenerate. Although a precise numerical estimate of such a separation depends on the parametrization of the Skyrme interaction, our calculations indicate that a relatively large, nearly 2 MeV separations are possible. The origin of the underlying mechanism, and the configurations that may produce such strong an effect, are discussed.

## ACKNOWLEDGMENTS

This research was supported in part by the Polish Committee for Scientific Research (KBN) under Contract No. 2 P03B 040 14, by the French-Polish integrated actions programme POLONIUM, and by the computational grants from the *Regionales Hochschulrechenzentrum Kaiserslautern* (RHRK) Germany, from the Interdisciplinary Centre for Mathematical and Computational Modeling (ICM) of the Warsaw University, and from the *Institut du Développement et de Ressources en Informatique Scientifique* (IDRIS) of CNRS, France (Project No. 960333).

FIGURES

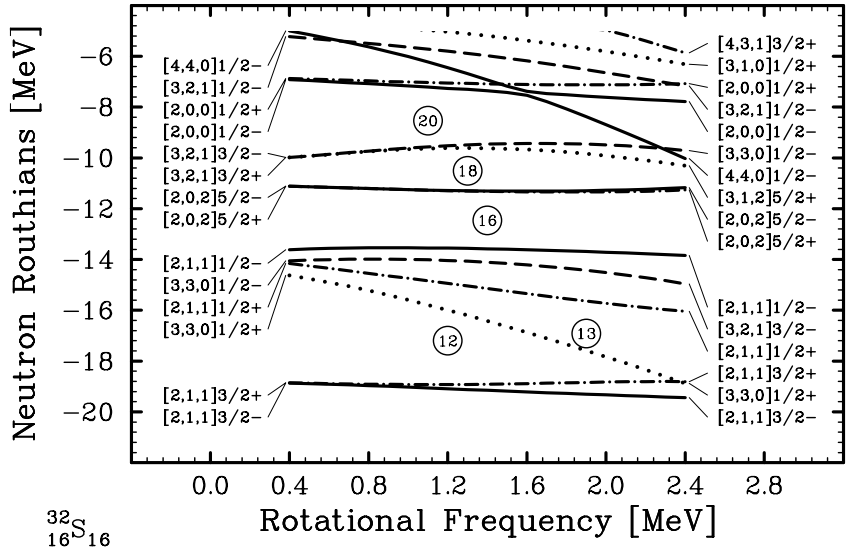

FIG. 1. Neutron single-particle routhians in the magic SD configuration of $^{32}$S calculated within the HF theory for the Skyrme SLy4 interaction. Lines denoting the four (parity, signature) combinations are: solid $(+,+i)$, dot-dashed $(+,-i)$, dotted $(-,+i)$, and dashed $(-,-i)$. Standard Nilsson labels are determined by finding the dominating HO components of the HF wave-functions at low (left set) and high (right set) rotational frequencies.

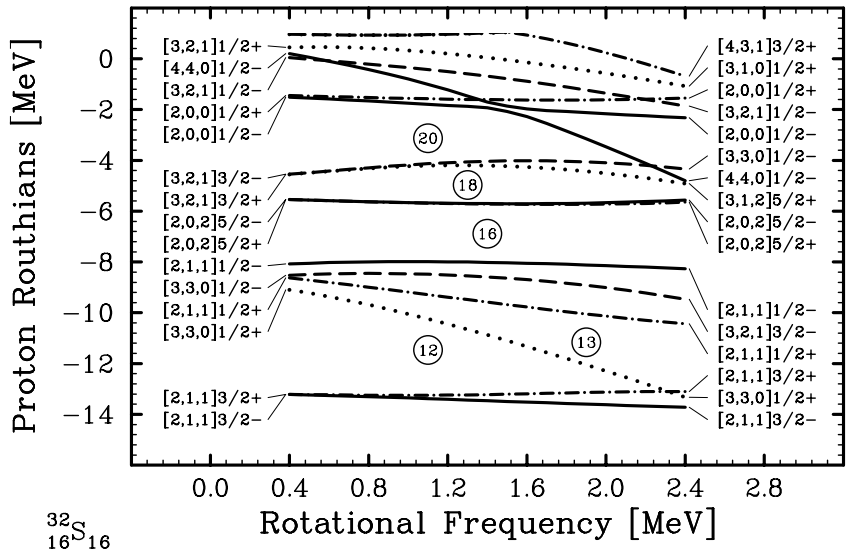

FIG. 2. Same as Fig. 1, but for the proton single-particle routhians.

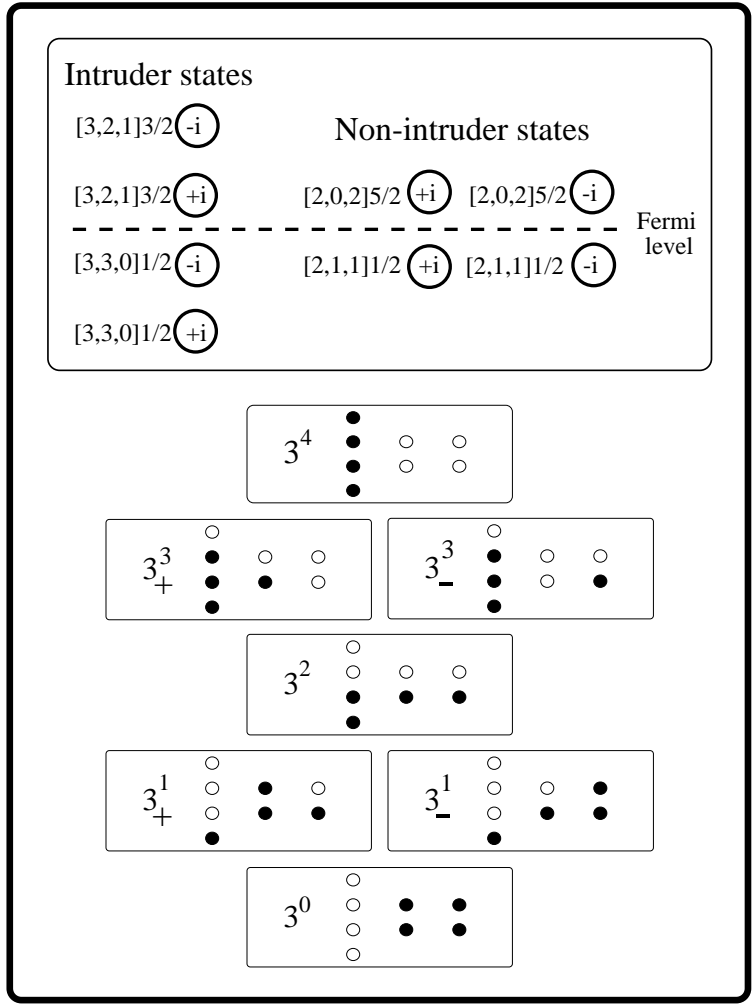

FIG. 3. Schematic diagram illustrating the single-particle neutron or proton orbitals (top), and the corresponding many-particle configurations (bottom), relevant for the description of SD bands in $^{32}$S. The top part gives the Nilsson labels and signatures ($r=\pm i$), inside the circles, of orbitals on both sides of the $N$=16 or $Z$=16 Fermi level. Four labels on the left-hand side represent the $N_0$=3 intruder states (negative parity), and four on the right-hand side represent positive-parity states. In the bottom part, the full circles stand for occupied, the open circles for empty states. Symbols $3^{n/p}$ give numbers $n$ or $p$ of (neutron or proton) occupied intruder states. Subscripts $\pm$ indicate whether the number of particles in the positive-parity $r=+i$ orbitals is larger than that in $r=-i$ orbitals, or vice versa.

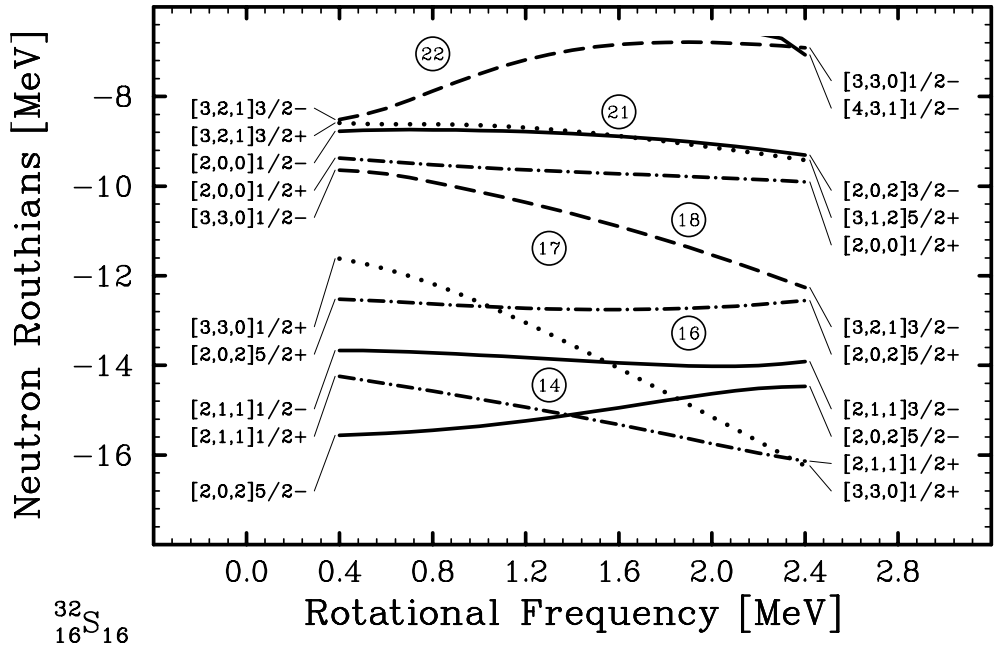

FIG. 4. Same as Fig. 1, but for the HF solution with the $3^1_- 3^1_-$ configuration.

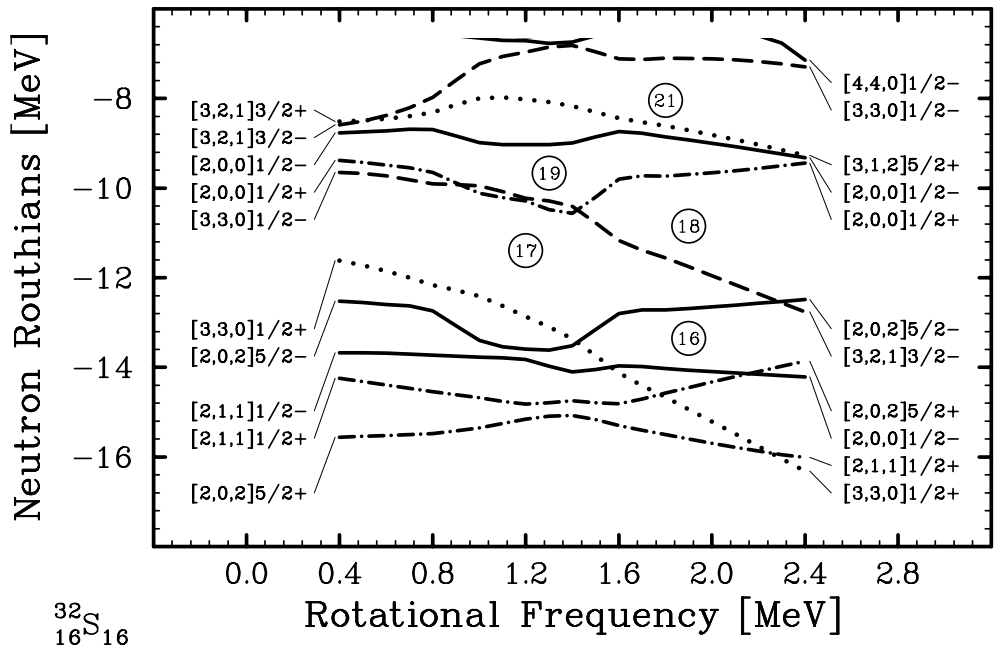

FIG. 5. Same as Fig. 1, but for the HF solution with the $3^1_+ 3^1_+$ configuration.

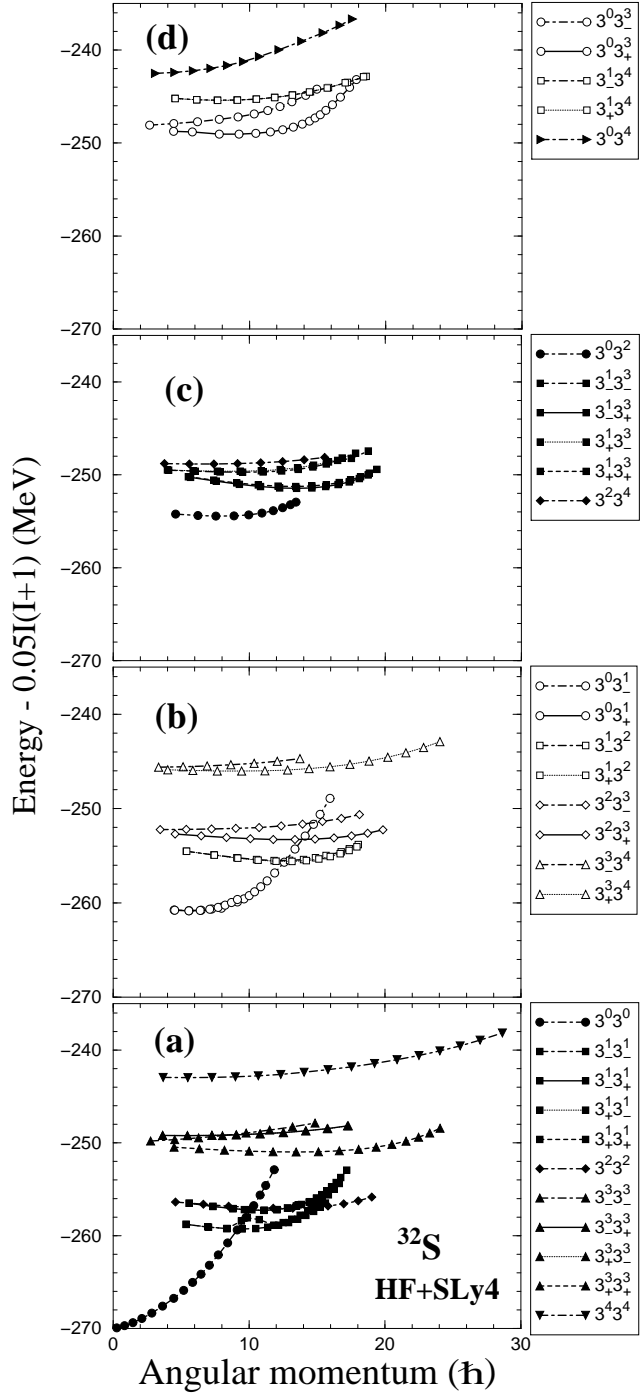

FIG. 6. Energies of the HF bands in $^{32}$S as functions of the angular momentum. A rigid-rotor reference energy of $0.05\,I(I+1)\,\text{MeV}$ has been subtracted to increase clarity of the plot. Full and open symbols represent the positive- ($\pi{=}+1$) and negative-parity ($\pi{=}-1$) bands. Long-short-dashed, solid, dotted, and dashed lines correspond to neutron ($r_n$) and proton ($r_p$) signatures being equal to, respectively, $(r_n,r_p)=(+,+), (+,-), (-,+)$, and $(-,-)$. Note that for even numbers of protons and neutrons the possible total signatures are $r_n = \pm 1$ and $r_p = \pm 1$; the latter should not be confused with the *single-nucleon* signatures taking the possible values of $\pm i$.

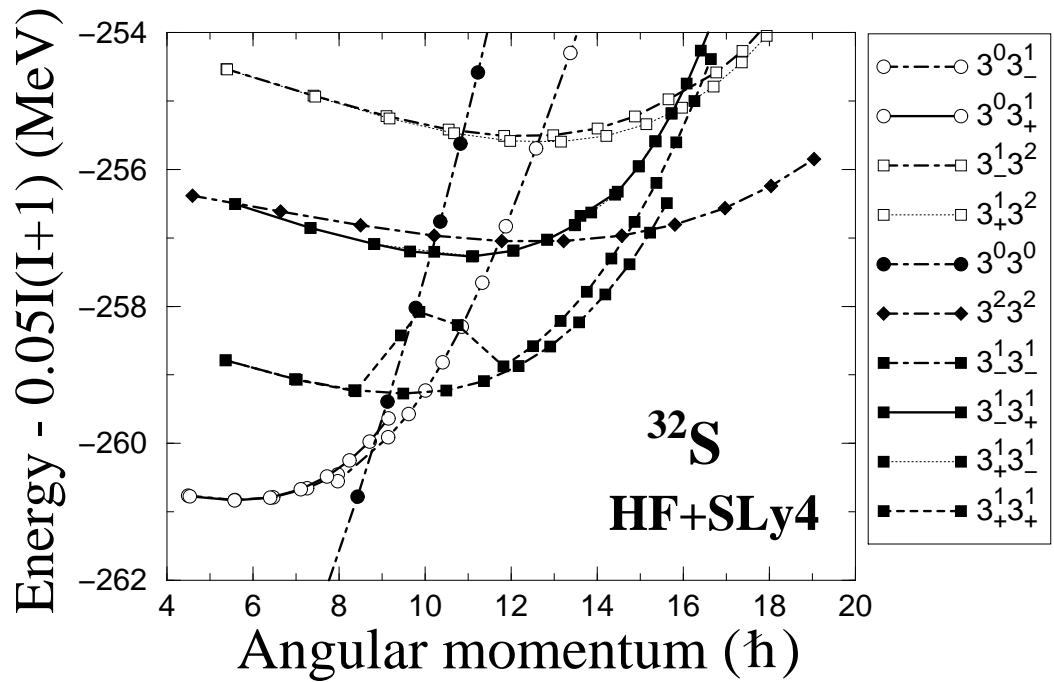

FIG. 7. Same as in Fig. 6, but for the yrast region of energies.

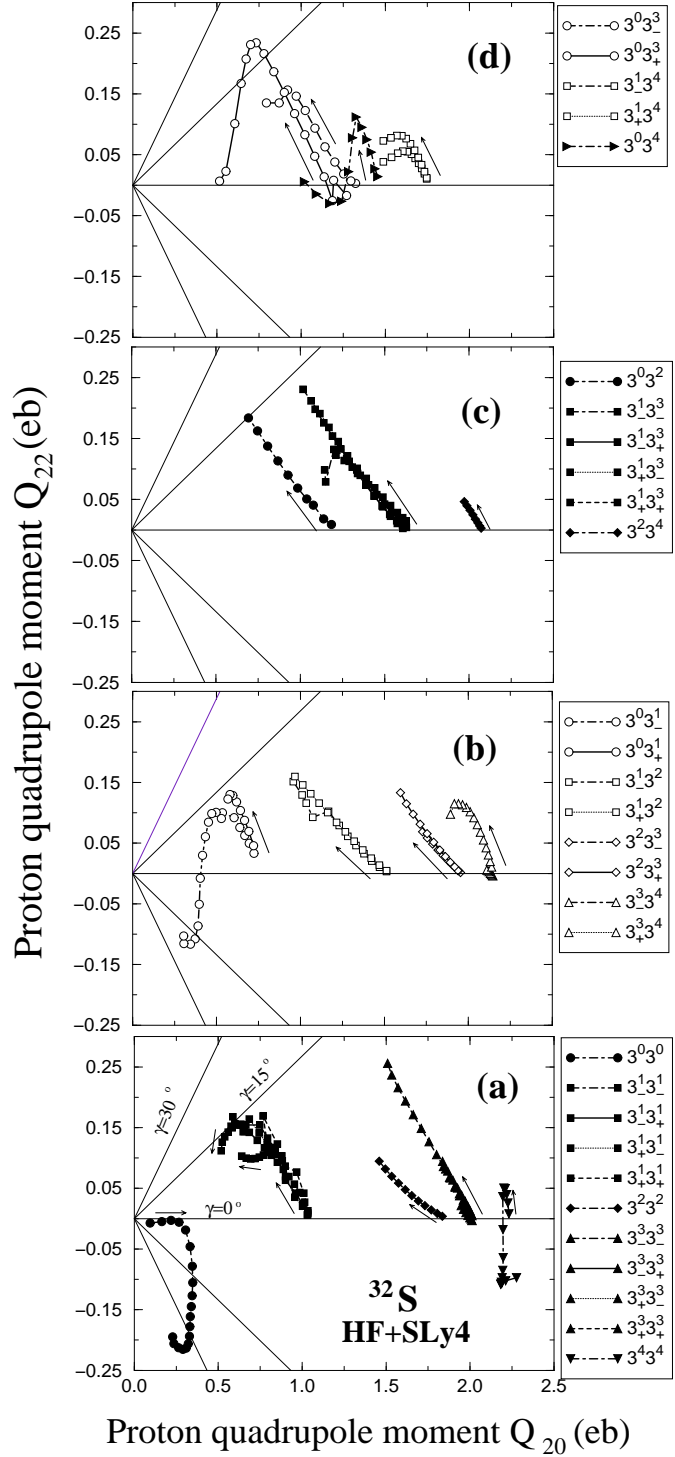

FIG. 8. Proton quadrupole moments of the HF bands in $^{32}$S, shown in the form of points on the $Q_{20}$–$Q_{22}$ plane. Since the variation of the multipole moments in function of the rotational frequency turns out to be regular the corresponding points form trajectories. Arrows indicate directions of increasing $\hbar\omega$. Note a large difference in scale between the $Q_{20}$ and $Q_{22}$ axes. The scales were adapted to the large differences between $|Q_{20}|$ and $|Q_{22}|$. The straight lines corresponding to $\gamma=\pm 15°$ and to $\gamma=\pm 30°$ have been drawn to facilitate reading the degree of non-axiality of the corresponding solutions.

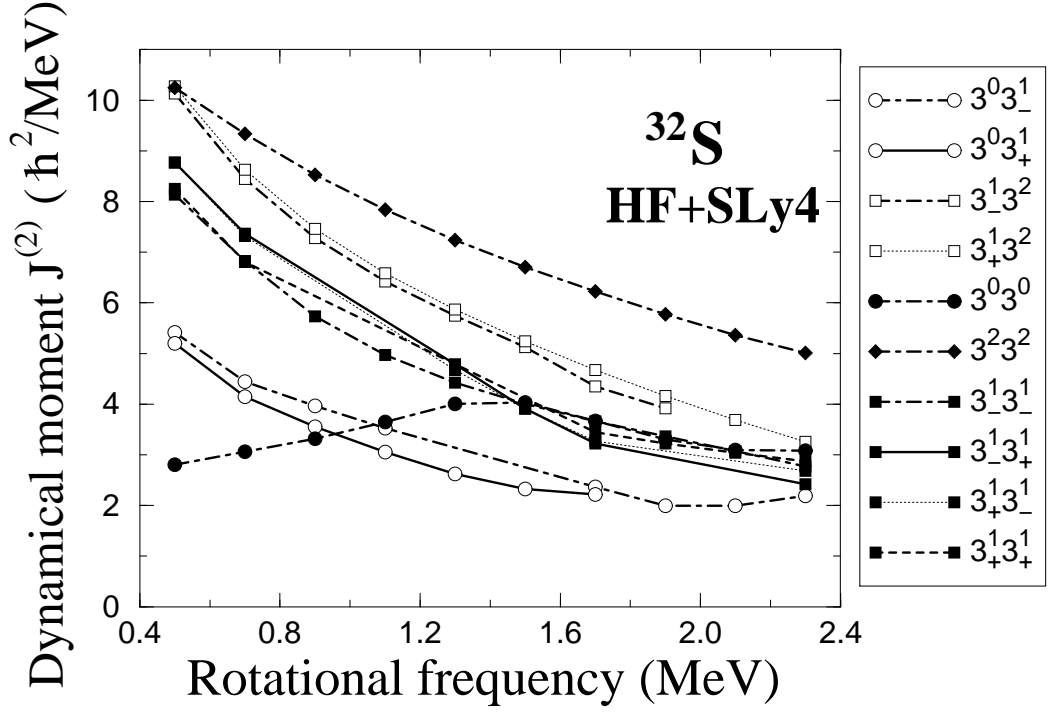

FIG. 9. Dynamical moments $\mathcal{J}^{(2)}$ of the HF bands in $^{32}$S as functions of the rotational frequency. The figure shows results for near-yrast bands selected in Fig. 7.

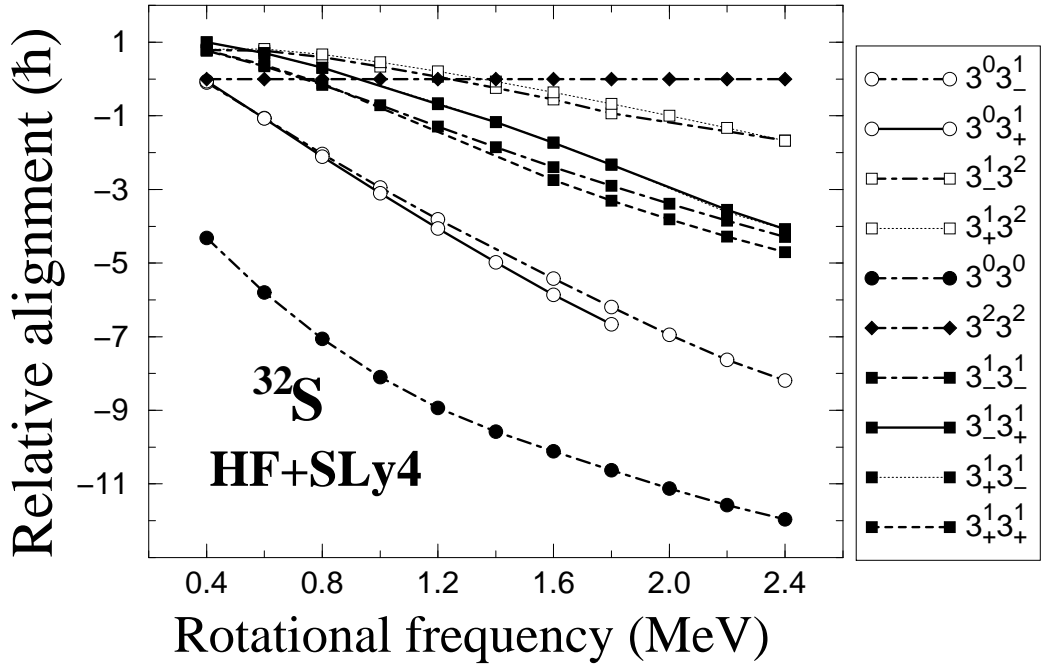

FIG. 10. Relative alignments $\delta I$ of the HF bands in $^{32}$S as functions of the rotational frequency, calculated with respect to the SD magic band in $^{32}$S. The figure shows results for near-yrast bands selected in Fig. 7.

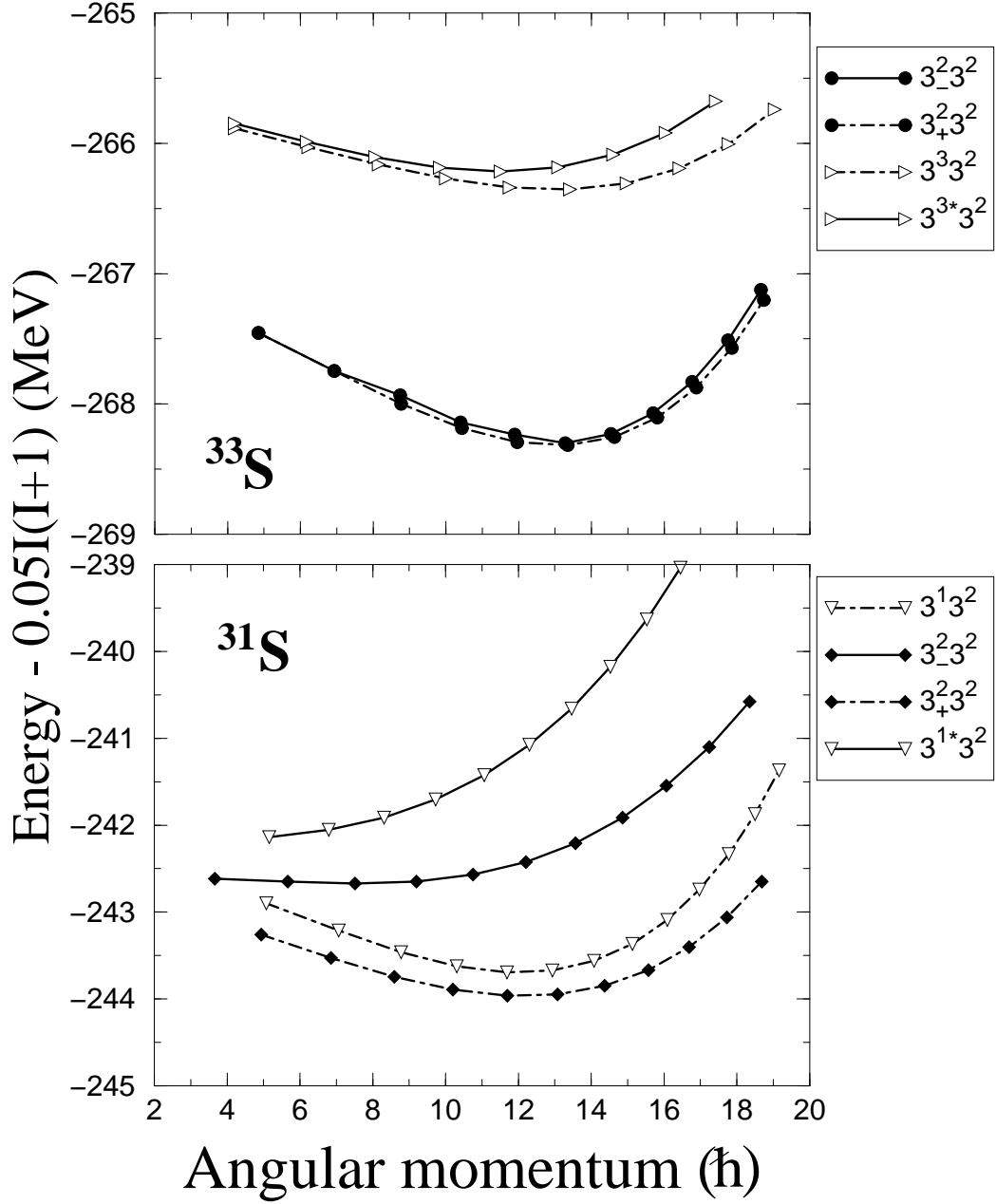

FIG. 11. Energies of the HF bands in $^{33}$S and $^{31}$S as functions of the angular momentum. A rigid-rotor reference energy of $0.05\,I(I+1)$ MeV has been subtracted to increase clarity of the plot. Full and open symbols represent the positive- ($\pi=+1$) and negative-parity ($\pi=-1$) bands. Long-short-dashed and solid lines denote signatures $r=+i$ and $r=-i$, respectively. Configurations $3^{1*}$ and $3^{3*}$ correspond to the highest negative-parity particles promoted to the next-to-lowest available intruder states. For $3^{1*}$, the first point corresponds to $\hbar\omega=0.6$ MeV.

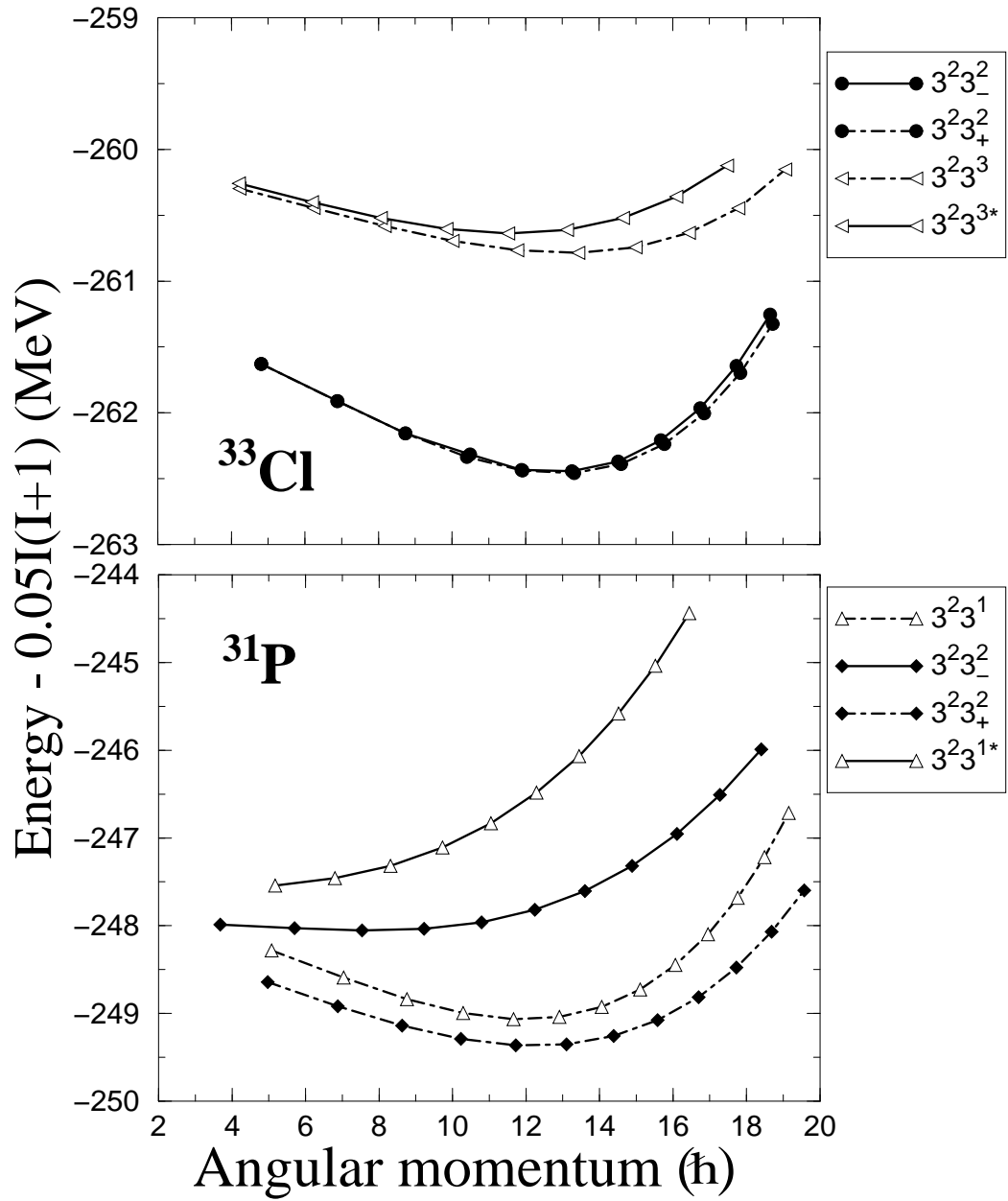

FIG. 12. Same as in Fig. 11, but for the $^{33}$Cl and $^{31}$P nuclei.

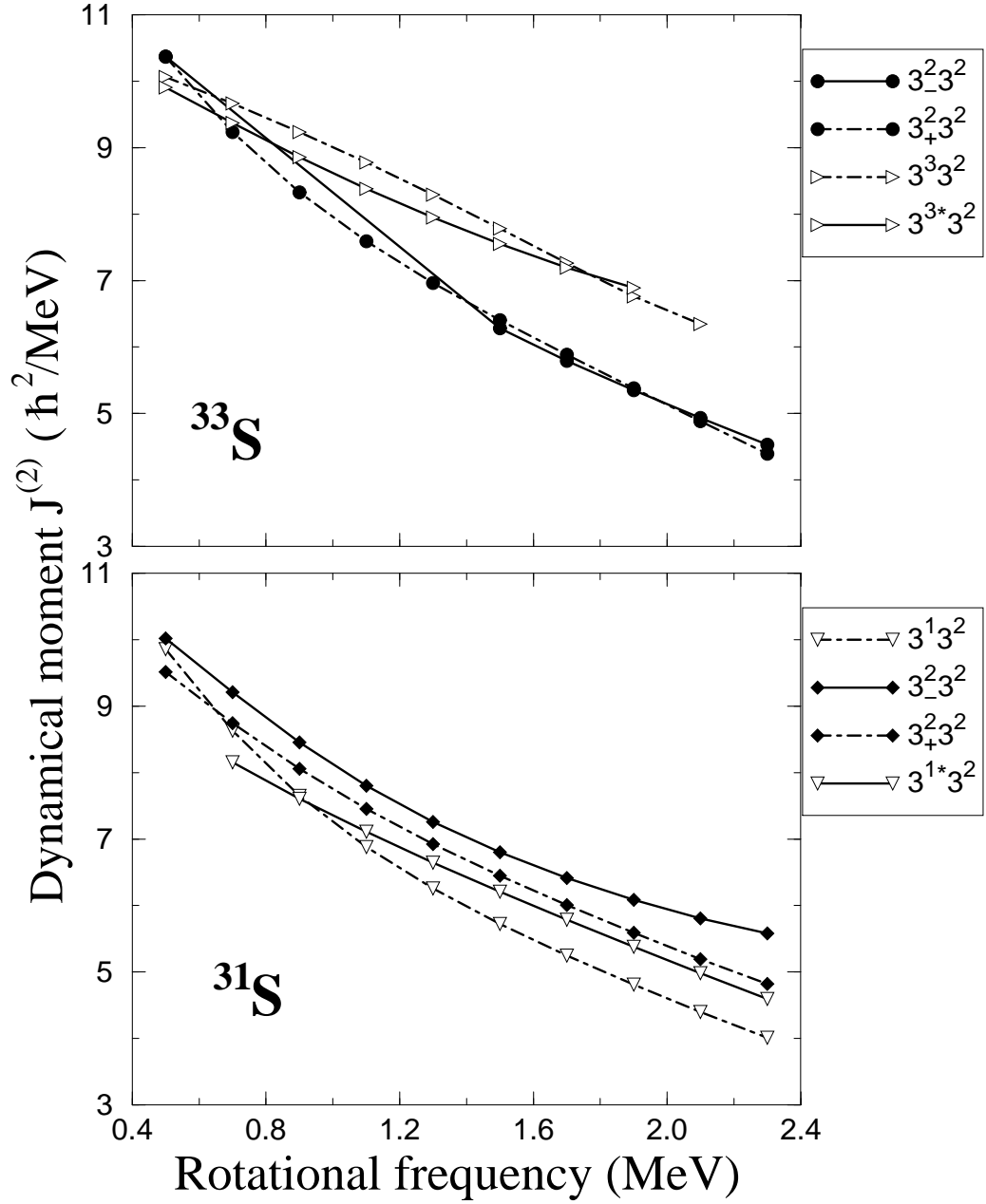

FIG. 13. Same as in Fig. 11, but for the dynamical moments $\mathcal{J}^{(2)}$.

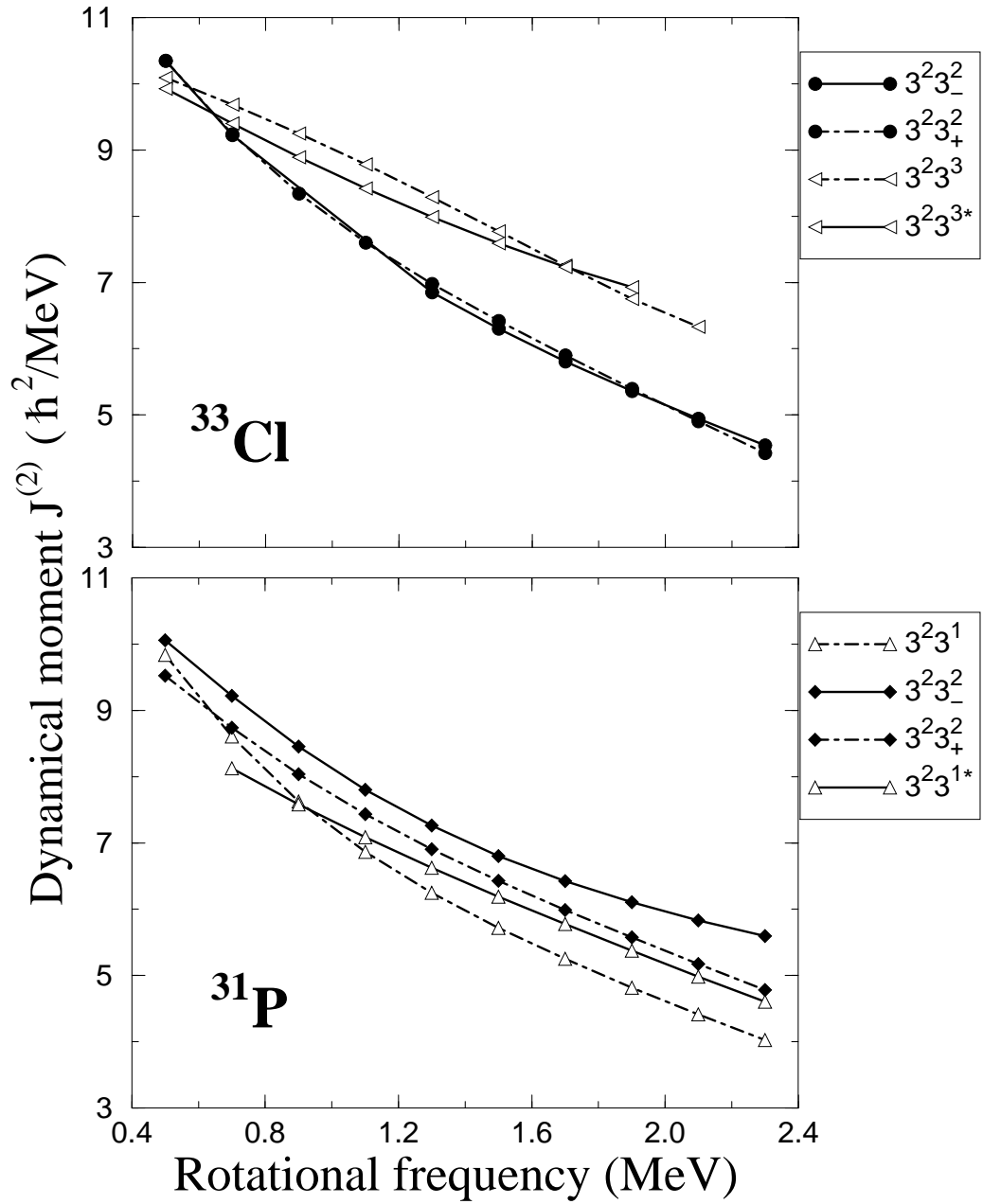

FIG. 14. Same as in Fig. 13, but for the $^{33}$Cl and $^{31}$P nuclei.

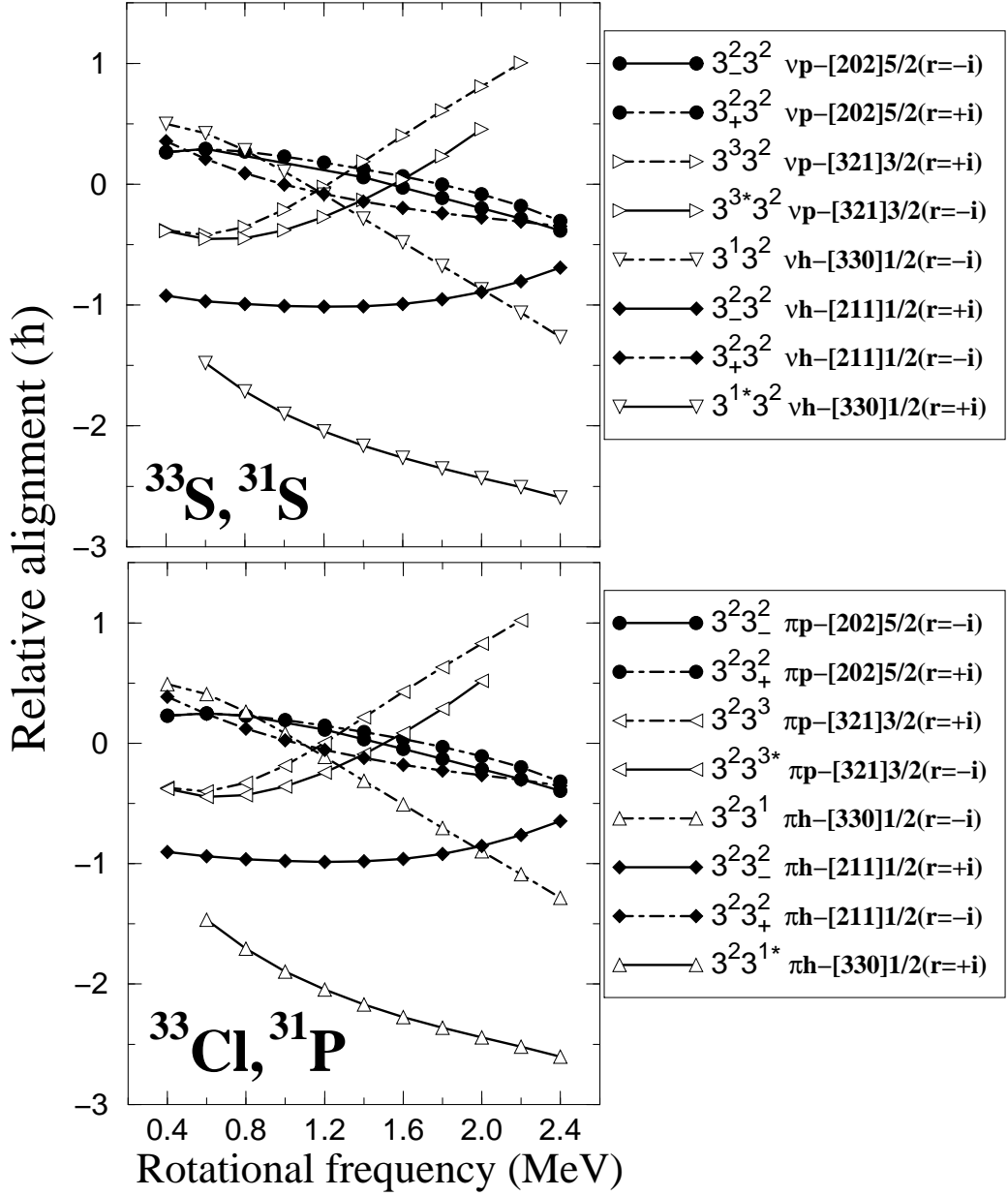

FIG. 15. Relative alignments $\delta I$ of the HF bands in $^{33}$S, $^{31}$S, $^{33}$Cl, and $^{31}$P as functions of the rotational frequency, calculated with respect to the SD magic band in $^{32}$S (see Fig. 11 for conventions used for symbols and lines). The Nilsson labels of particle (p) or hole (h) orbitals, which make the difference between the given band and the magic band in $^{32}$S, are indicated on the right-hand side.

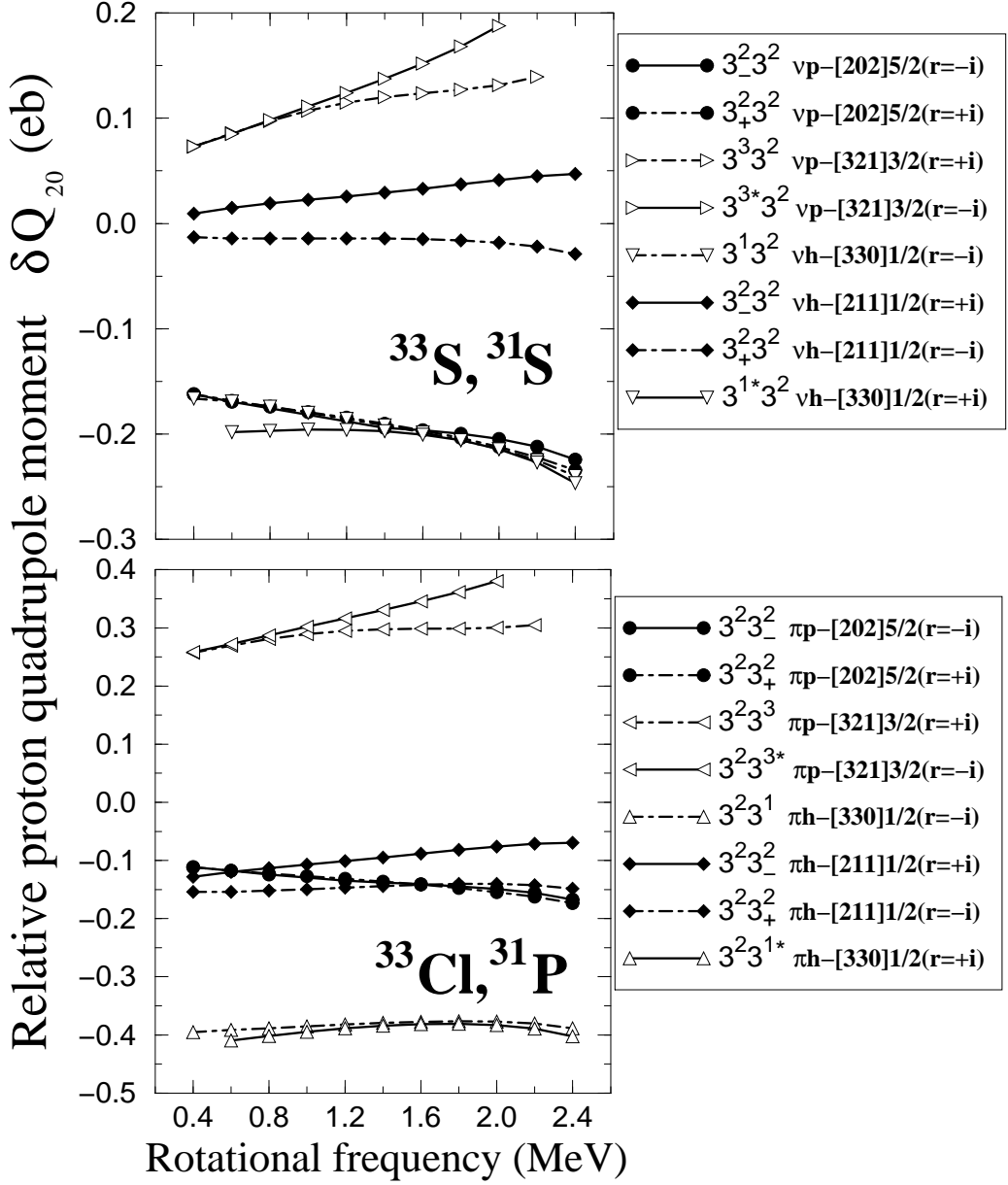

FIG. 16. Same as in Fig. 15, but for the relative proton quadrupole moments $\delta Q_{20}$.